\documentclass{aa}  
\usepackage{graphicx}
\usepackage{txfonts}
\begin{document}
\newcommand{\kms}{{\, \rm km~s}^{-1}}
\newcommand{\obj}{HE~0450--2958}
\title{Slit and integral-field optical spectroscopy of the enigmatic quasar \obj.
\thanks{based on observations made with VLT/UT1  (program 76.B-0693(B)),
and with the VLT/UT3 (program 72.B-0268(B)) at ESO-Paranal observatory 
in Chile, in addition with NASA/ESA Hubble Space Telescope observations from Cycle 13, proposal \#10238.}
}

   \author{G. Letawe\inst{1}, P. Magain\inst{1} \and F. Courbin \inst{2}}

   \offprints{G. Letawe}
\institute{Institut d'Astrophysique et G\'eophysique, 
 Universit\'e de Li\`ege, All\'ee du 6 Ao\^ut, 17 Bat B5C, B-4000 Li\`ege, Belgium\\
 \email{gletawe@ulg.ac.be}
 \and
 Laboratoire d'Astrophysique, Ecole Polytechnique F\'ed\'erale de Lausanne (EPFL),  Observatoire, CH-1290 Sauverny, Switzerland \\
             }

   \date{Received ; accepted}

 
  \abstract
   {Interest in the quasar HE0450-2958 arose following the publication of the non-detection of its expected massive host, leading to various interpretations.}
   {This article investigates the gaseous and stellar contents of the system through  additional VLT/FORS slit spectra and integral field spectroscopy from VLT/VIMOS.}
   {We apply our MCS deconvolution algorithm on slit spectra for the separation of the QSO and diffuse components, and develop a  new method to remove the point sources in Integral Field Spectra, allowing extraction of velocity maps, narrow-line images, spatially resolved spectra or ionization diagrams of the surroundings of \obj.}
   {The whole system is embedded in gas, mostly ionized by the QSO radiation field and shocks associated with radio jets.   The observed gas and star dynamics are unrelated, revealing a strongly perturbed system.  Despite   longer spectroscopic observations, the host galaxy remains undetected.}
   {}

               
\titlerunning{HE0450-2958: 2D and 3D spectroscopy}
\authorrunning{G. Letawe et al.}
   \maketitle
%

\section{Introduction}

In the framework  of a systematic spectroscopic  study of  20 QSO host
galaxies  at   low  redshift   (\cite{letawe07}),  we     discovered a
particularly  interesting   object  with no detected host  galaxy:
\obj. While all the  other QSOs in the   sample show a host  galaxy
even in the short acquisition  exposures taken  with  the ESO Very Large
Telescope (VLT),  that of \obj\ remains  undetected, both on  our deep
VLT optical spectra  and on high  resolution ACS images taken with
the Hubble Space Telescope (HST). From these observations Magain et al. (2005, hereafter M05) set a deep upper limit,  implying that, given the luminosity  of  the QSO, the host galaxy is underluminous by at least a factor of six.

\begin{figure}[h!]
\centering
\includegraphics[width=6.5cm]{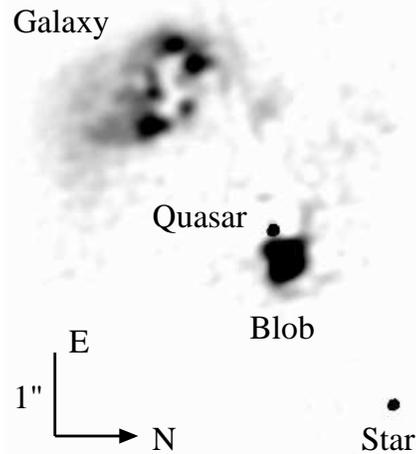}
\caption{HST/ACS image of \obj, obtained with the High Resolution Channel
of  the instrument,  through the  F606W  filter.   The  image has been
spatially deconvoled using the MCS algorithm (adapted from
M05). }
\label{imgHST}
\end{figure}

\obj, at $z=0.285$,  lies in  a rather complex environment, as shown in 
Fig.~\ref{imgHST}.  Aside from a foreground star  which happens to be close to
the line of sight,  about 1.7\arcsec\, to  the North-West of the  QSO,
two other objects are at the  same redshift as the  QSO: 1- a strongly
distorted  companion galaxy, located at   1.5\arcsec\,(6.5 kpc) to the
South-East of   the QSO,  2-  a  compact   blob whose  spectrum  shows
prominent emission  lines, typical for highly  ionized gas, but with no
trace  of continuum light.   This gas is  most probably ionized by the
QSO  itself, as   suggested    by the  emission   lines    flux  ratios
(M05). 

Since the publication  by Magain et al. (M05), several papers  have proposed
alternative explanations  to the fact that no  stellar light is detected
around \obj.  In particular, the QSO could have  been ejected during a
3-body  interaction,  where a binary  black  hole kicked the  QSO off
during a merger event (\cite{hoff}). An alternative ejection mechanism would be recoil due to  gravitational
wave emission (\cite{haehnelt}).  Although this  scenario is supported
by  the recent  discovery of   a  genuine triple QSO   (\cite{djo07}),
further simulations tend to weaken these hypotheses (\cite{merrit06}).
In  their analysis  Merritt et al. note  that \obj\, emission  lines are
narrower than for the majority of Type I QSOs.  They suggest  \obj\ to be a  high luminosity analog  to  the  so-called narrow-line Seyfert   1
galaxies (NLS1s), where a lower-mass supermassive black hole (SMBH)
accretes matter at a high rate, possibly above the Eddington limit. This
is indeed supported  by X-ray analyses   (Zhou et al.  2007).  If  the empirical relation between the mass of  the SMBH and the mass of
the host galaxy bulge holds  (e.g. Marconi \&   Hunt 2003), one  would
then expect a  lower luminosity  host  galaxy for \obj,  possibly just
below the detection limit of M05.  Estimates of the host magnitude upper limits were also performed using a different method by Kim et al. (2007), confirming the value of M05.  These authors argue, in agreement with Merritt et al. (2006), that the bulge might be just too faint for detection with the observations made so far.

Other  possible explanations  for   the non-detection of the  host  are
internal extinction  by dust  and/or  the absence of  a  genuine young
stellar population.  In both cases, the host galaxy might be at the limit of detection of the M05 ACS images.
 To date, the only evidence suggesting the presence of a host comes from the radio observations of \cite{klam07}. These authors point out that the observed fluxes in different radio bands are consistent with star formation-induced emission and, moreover, that they obey the Far Infrared/Radio Continuum correlation found for star forming galaxies.

The  present  paper focuses   on new  long  slit  and   integral field
spectroscopy carried out  at the ESO-VLT,  with the aim of  characterizing
the physical state   of the gas  surrounding \obj\   and also to
attempt to detect the  stellar continuum  of the  host  galaxy, if  it
exists.


\section{Observations and reductions}

\subsection{Slit (2D) spectroscopy with VLT/FORS2 MXU}

Our new multi-slit  observations of \obj\ were  obtained with  the ESO-VLT
and the FORS2 instrument  in  the MXU (mask  exchange unit)  mode.  We
used  the  G600RI grism,  covering the  spectral   range from 5145 to
8470\AA, i.e., the  major stellar  and gas  emission lines in the redshifted spectrum of
\obj.  The observations were taken between February and April 2006, as
5 sets of 6 different exposures,  for a total  exposure time of 3.75 hours
in dark time. Their spectral resolving power is R=1000.

The slitlets of the MXU mask are all 1\arcsec\ wide, but have variable
lengths.  As  in M05  and Letawe et al. (2007)  we place
slitlets on  the quasar  and on several  surrounding  stars of similar
magnitude.    The spectra of the   stars are used as   PSF templates to remove the  quasar
light using the version of the MCS algorithm (\cite{MCS}) for spatial  deconvolution
of spectra (\cite{mcs_2}).   The orientation of  the slits is selected
to allow the simultaneous observation of  the
quasar, the  companion galaxy and the foreground star.  The  remaining  slits  are used to
observe the surrounding  galaxies.  The seeing during the observations
varied from 0.5 to 1.3\arcsec.

The basic  reductions are performed  with \textsc{iraf} tasks, leading
to flatfielded, wavelength calibrated and sky subtracted spectra, with
1.5 \AA/pixel in the spectral  direction and 0.126\arcsec/pixel in the
spatial direction.

\begin{figure}[t!]
\centering
\includegraphics[width=8.cm]{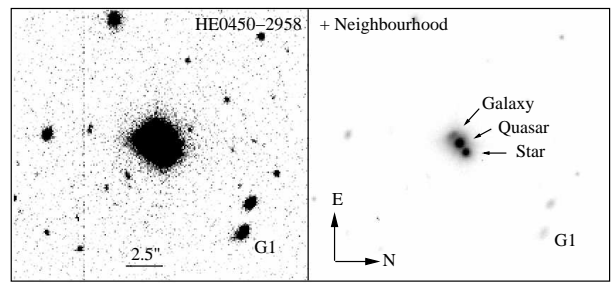}
\caption{Immediate surrounding of \obj. The image is the combination 
of the 30-second exposures in  $B$,  $V$  and $R$ obtained at the 
VLT/FORS2 to design the slit masks. Two different 
intensity  scales are shown to display the full dynamical 
range of the image. }
\label{preim}
\end{figure}

In addition to the slit spectra, several images of the field were
taken with FORS2 to  design the slit masks.  Three 30-seconds
exposures were  taken for this purpose, through  the  $B$, $V$ and $R$
filters.   Figure~\ref{preim}   shows the   combination of  these  three
exposures, with  two different   intensity contrasts to  enhance  both
faint and bright structures.  They reveal  many faint objects within a
few arcseconds of \obj. However, as  the quasar and most point sources
of similar  brightness are saturated, no  further data processing is
carried out on these images.

\begin{figure}[t!]
\centering
\includegraphics[width=8.8cm]{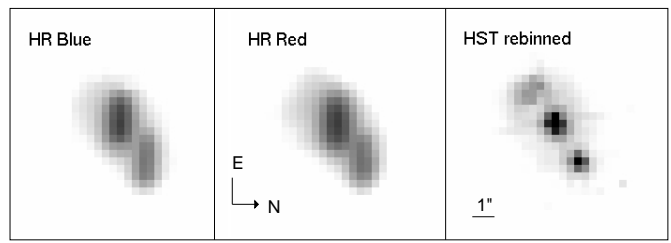}
\caption{VIMOS reduced data integrated over the whole spectral range, for 
the HRblue (left)   and HRred grisms  (middle). For  comparison we
show the HST/ACS    image  on the   right  panel, rebinned    at the same
orientation  and  resolution as the  VIMOS  data. Intensity scales are
logarithmic.}
\label{rawimg}
\end{figure}

\subsection{Integral field spectroscopy (3D) with VLT/VIMOS}

The integral field   unit (IFU) of  the VIMOS  instrument was used  to
obtain 3D   spectra of \obj.    In this  configuration,  a spectrum is
obtained for each position  on a $13\arcsec\times 13\arcsec$ field  of
view, every 0.33\arcsec. The spectral resolution is R$\sim$2500.

Two  settings (HRblue and HRred) are  used to cover the wavelength   range  $4200\ \AA <
\lambda  < 8600\ \AA$.  Unfortunately,  the gap  in wavelength between
the two grisms (6150 to 6350\AA)  contains the H$\beta$ emission line,
at the  redshift ($z=0.285$) of  HE0450-2958.  The spectra provided by
the  optical fibers  fed   by the telescope    are organized in   four
quadrants containing $20\times20$   fibers   each. 

The observations consist  of  $7\times 40$ min exposures with the  blue grism in
December  2003, and  of $11\times  19$ min with  the  red  grism, taken
between December 2003 and February  2004.  The seeing varied from 0.45
to 0.9\arcsec in the blue  (0.63\arcsec on average)  and from 0.53  to
1.2\arcsec\, in the red grism (0.76\arcsec on average).

 At the    time  of observation, the HRred grism was not  available on the fourth quadrant
and was replaced  by HRorange, covering  a bluer wavelength  range and
with a different response curve.  This complicates the analysis of the
fourth quadrant. As a consequence, only 7 out of the 11 HRred exposures were
useful due to  problem in the matching of   HRred and HRorange  on the
edges of  the  fourth  quadrant, located at the NW corner of the field (see Fig. \ref{rawimg}).  In this figure,  we show  the VIMOS
field of view in the  two settings and  we compare it with the HST/ACS
image degraded to  the resolution and pixel size  of  VIMOS. The VIMOS
PSF is  highly elongated.  Since our   reduction procedure (see below)
corrects  for atmospheric refraction,  this elongation is probably
due to a tracking problem.

The   observations are reduced using  various  pipelines and routines,
taking advantage  of the specific capabilities   of each package. The main
steps of the reduction can be summarized as follows:

\begin{itemize}

\item{The cosmic-ray rejection is carried out on the raw data by applying
the {\tt filter/cosmic} task of \textsc{midas} to the 2D spectra  in each  
of the four quadrants.}

\item The bias subtraction, flatfielding, wavelength calibration, extraction
of the spectra for all fibers, and  flux calibration are done using
the \textsc{esorex} recipes   from  the  VIMOS  Pipeline  provided  by
ESO. The reduced data with the HRblue grism have a spectral sampling of
0.54\AA/pixel  and   the   data  taken   with the   HRred grism   have
0.58\AA/pixel. 

\item A basic 2D sky subtraction is carried out with the {\tt background} \textsc{iraf} task, by linear interpolation on  each spectral element at positions devoid of sources.

\item Whenever a strong sky emission line is not removed well enough by the \textsc{iraf} task,
a \textsc{fortran} routine is applied to clean the spectrum by linear interpolation
between the regions unaffected by these sky emission lines.

\item \textsc{python} routines are then used to re-organize the spectra 
into  3D data cubes of 40  x 40 spatial resolution elements each containing
 a 1D spectrum. We use the word ``spaxels'' in the following,
when referring to each of  the  $40 \times  40$  elements, to
avoid confusion with pixels.

\item The reduced exposures are then aligned using a \textsc{python} task,
which shifts the spatial location of the quasar for each spectral element
to the center of the IFU field. This ensures that atmospheric refraction is 
corrected in each exposure.

\item The aligned exposures are co-added with appropriate weighting, taking into account the different coverage of the sky due to offsets between exposures.

\item The resulting combined data cubes are smoothed using a Gaussian 
kernel  of 2.5\AA \, full width at half maximum (FWHM). 

\end{itemize}


\section{Slit spectroscopy}

Because of the very high contrast between the QSO and 
its putative host galaxy, particular care must be paid to the 
decomposition of the data into  point-like and  extended 
components, both in the slit and IFU spectra.

\begin{figure}[t!]
\centering
\includegraphics[width=8.7cm]{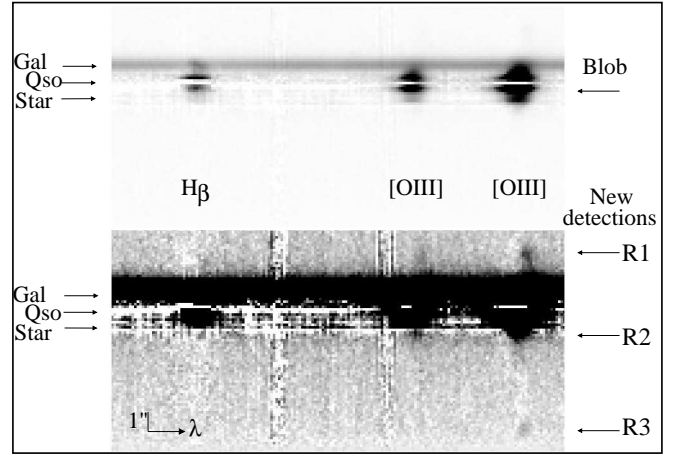}
\caption{Combined FORS2 spectrum of \obj, in the region where the 
deconvolution shows acceptable residuals. The point sources (QSO+star)
have been removed. The spectra are shown with different intensity scales
in the  two panels.  The 3 new emission line regions are indicated. The
source of their ionization is the QSO radiation.}
\label{deco3}
\end{figure}

\begin{figure}[t!]
\centering
\includegraphics[width=9.3cm,height=6.7cm]{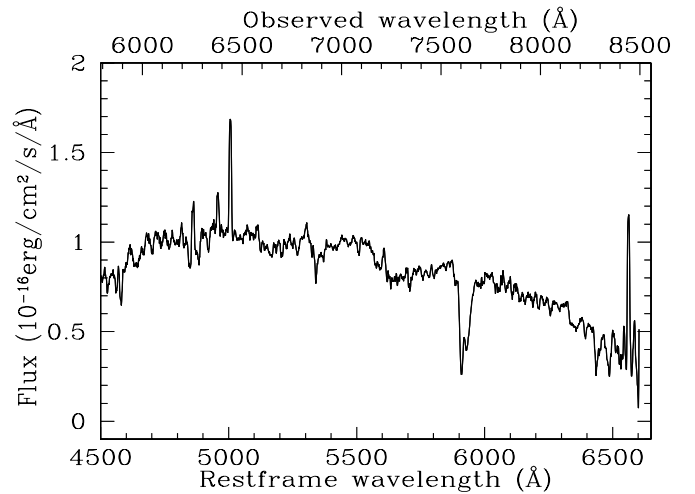}
\caption{FORS2 spectrum of the late-type 
galaxy G1 shown in Fig.~\ref{preim}. The measured redshift is $z=0.2866$}
\label{neigh}
\end{figure}

The   FORS2/MXU spectra of  \obj\  are all  deconvolved  using the MCS
algorithm for  slit  spectra  (\cite{MCS}; \cite{mcs_2}), allowing   a
separation  of   the  QSO   light  from  the   underlying  extended
objects.  This method has  been  used successfully on  many occasions,
either on data of QSO host galaxies (\cite{courb02};
\cite{letawe03}; M05; \cite{letawe07}) or on data of gravitationally 
lensed QSOs (e.g., \cite{alex06}; \cite{alex07}).  

The  goal   of  the present  observations   was to   go  deeper than in M05,  to attempt the detection of a stellar  continuum in the putative
host galaxy and to determine the ionization state  of the gas far away
from the  QSO. Unfortunately, optimal  separation of the QSO and
host cannot be achieved in this case:  the seeing during  the observations, worse than
originally planned, is of the  same order as  the slit width, resulting
in a  loss of information on  the  wings of the  PSF.  This  effect is
enhanced by  the   combined effects of a slight  slit  misalignment with
respect  to  the   point  sources, and   of  differential  atmospheric
refraction.  The consequence is that the  QSO and the PSF stars are
clipped  in  different ways, making it  impossible  to  build reliable
PSFs.  In addition, the  overall quality of  the PSF does not allow simulations  to be performed (see M05) in  order to place a useful
limit on the magnitude of the continuum.

We  show  in Fig.~\ref{deco3}  a representative  portion  of  the  data,  between H$\beta$ and [OIII],  where the deconvolution process is acceptable, i.e.,  a region where  the remaining flux after removal of the point sources is non-negative, and where the residuals of the deconvolution process do not present obvious features of a misadapted PSF.  The image is the  combination of all  5 of our individual, deconvolved,  spectra.  Even with bad
deconvolution   residuals,  as shown  by  the consequent noise  at the
position of  the   point  sources, and thanks to the deep exposures, three new emission-line    regions  are
detected  at  large  distances from the  QSO (see Fig.~\ref{deco3}).

The  three regions R1, R2,  and R3, are located  at  2.6, 1.5 and 5.9 arcseconds from the QSO. This corresponds to projected distances of 12, 7 and 27 kpc respectively, at the redshift of \obj.  While regions R1 and R3  had not been detected previously, region R2 was  reported in \cite{klam07}, on the basis  on our FORS1/MOS analysis (M05). It is almost superimposed  on  the foreground star lying  to the  North-West of the  QSO.    

\begin{figure*}[th!]
\centering
\includegraphics[width=18.0cm]{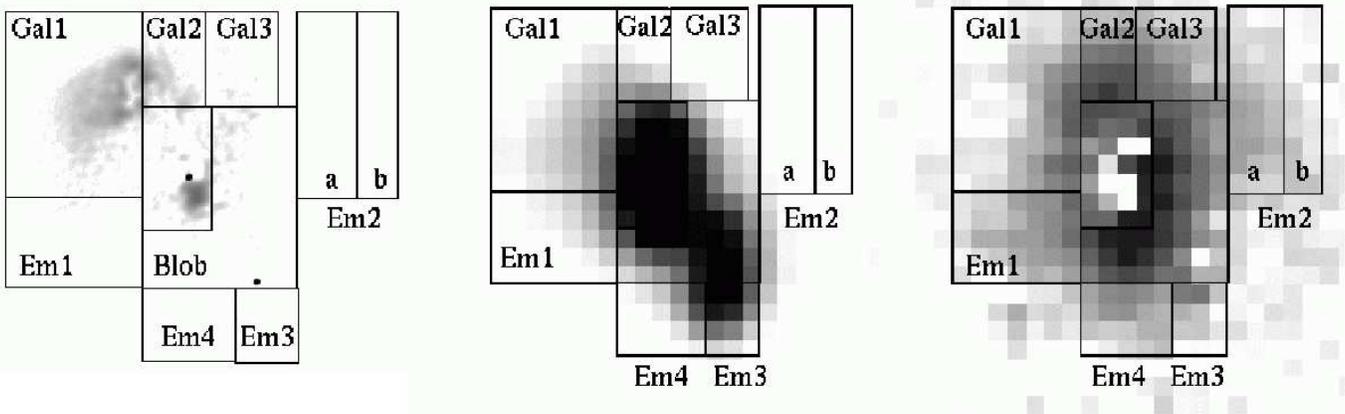}
\caption{Spatial regions over which we integrate individual 1D spectra. The 
regions are shown  in overlay on  the  deconvolved HST/ACS  image (left
panel), and on the VIMOS image reconstructed from the full datacube by
integrating over the full  wavelength range (middle panel). The  image
on the right is the result of the subtraction of the QSO and of nearby
star spectra. In this case, the wavelength range  around the redshifted [OII] emission  line has been  selected.  The central 30 spaxels of the QSO, very noisy
after subtraction, are not used when integrating the signal over the regions.}
\label{zones}
\end{figure*}


Only emission lines are seen in  the 3 regions,  despite  the depth
of the observations.  In the  3 cases, the  [OIII] doublet is detected
while the H$\beta$  line is absent.  For region R1, this converts into a
lower   limit     of [OIII]/H$\beta>$ 7.7,     and   for   region   R2,
[OIII]/H$\beta>$ 28.8.  The  third region is so    faint that the lower
limit  derived is  irrelevant.  Following  the ionization diagrams  of
\cite{veilleux},    a  ratio  above  6  can   safely  be associated with
ionization by QSO   radiation. Our FORS2 observations therefore indicate
that the source  of ionization in  these regions is the  QSO itself and, as a consequence, that ionizing photons can escape the QSO surroundings to  ionize gas clouds as far as at least 7 and 12 kpc. This gives further support to the idea that, if dust is present in the immediate surroundings of the QSO, its geometry is such that UV radiation can escape freely in several directions: towards us (no detected reddening of the QSO and blob spectra, M05) and towards these emission line regions.

There is still no detection of a host galaxy in these spectra but, as previously mentioned, the quality of the deconvolution process is not optimal. However, a comparison of the flux in the companion galaxy and the remaining flux under the QSO (i.e. the putative host) reveals that the host should at least be two magnitude fainter than the companion galaxy, in the wavelength range between [OIII] and H$\alpha$ (equivalent to the observed filter R). This would give a magnitude for the host of M$_V\sim-$21, compatible with M05 and Kim et al.\  (2007) estimates. Given that most of the flux detected under the QSO is likely to be due to deconvolution artefacts caused by an inaccurate PSF, the putative host might be much fainter than that upper limit.


\begin{figure*}[p]
\centering
\includegraphics[width=17cm,height=7cm]{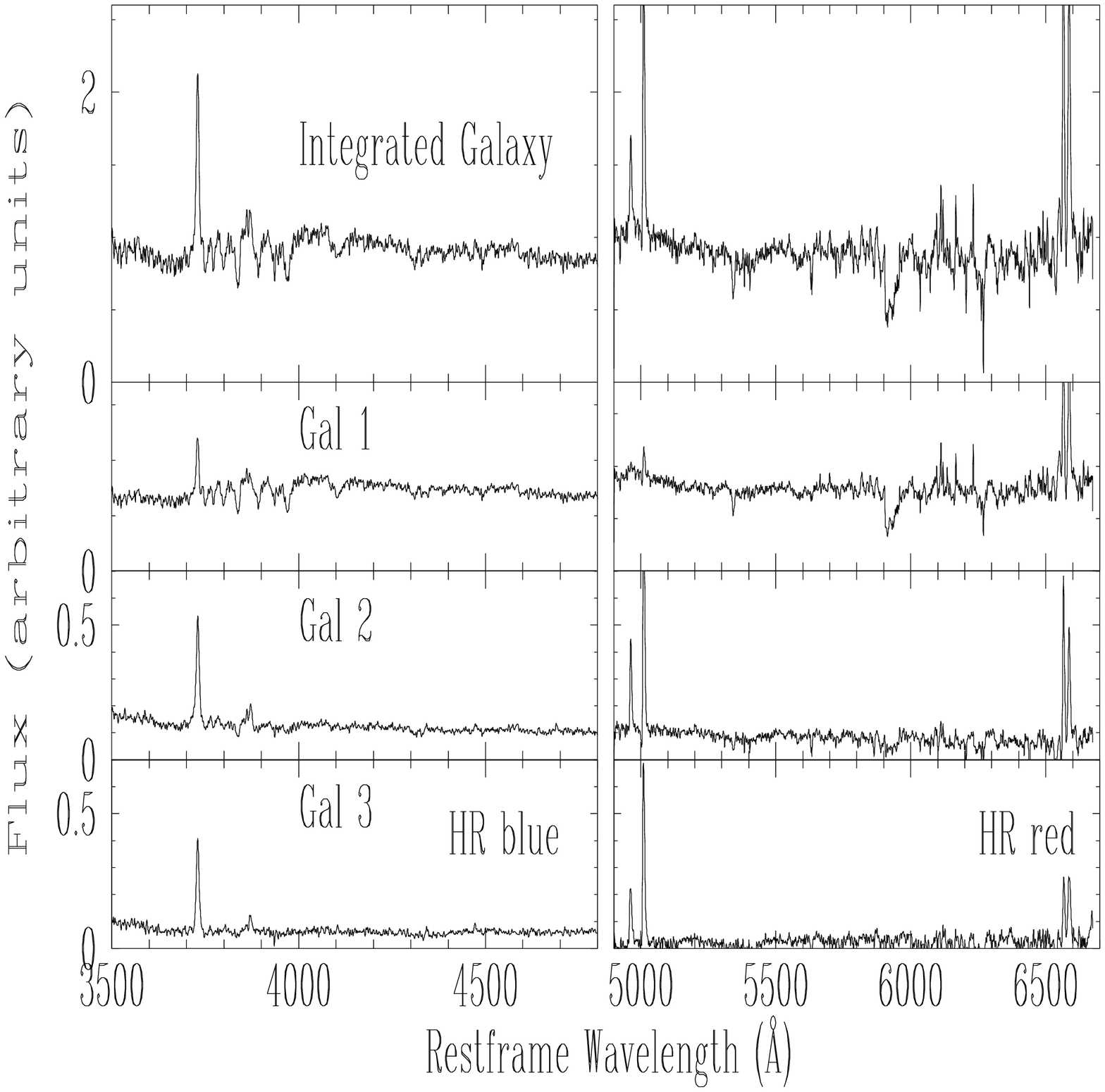}
\caption{Spectra of the companion galaxy in the HRblue and HRred VIMOS grisms.
For the two settings, the spectrum in the top panel is the sum of the spectra for
the 3 regions  labeled  Gal1  to      Gal3   in
Fig.~\ref{zones}, and shown in the lower panels. }
\label{gal}
\vskip 15pt
\includegraphics[width=17cm,height=7cm]{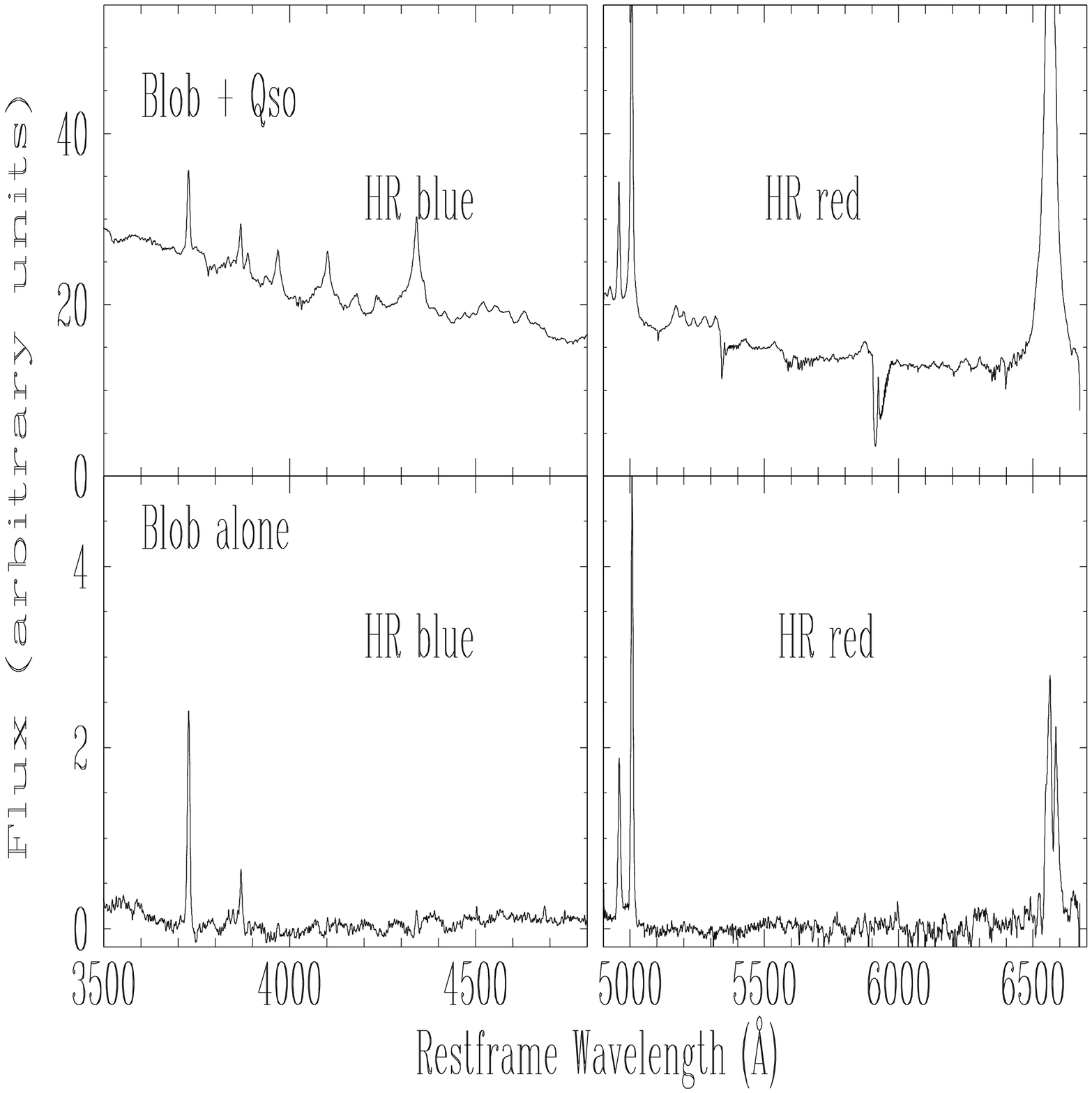}
\caption{Spectra corresponding to the ``blob'' region in the HRblue and HRred grisms.
The spectra are shown before and after the QSO is removed, in the upper
and lower panels respectively .}
\label{blob}
\vskip 15pt
\includegraphics[width=16cm,height=7cm]{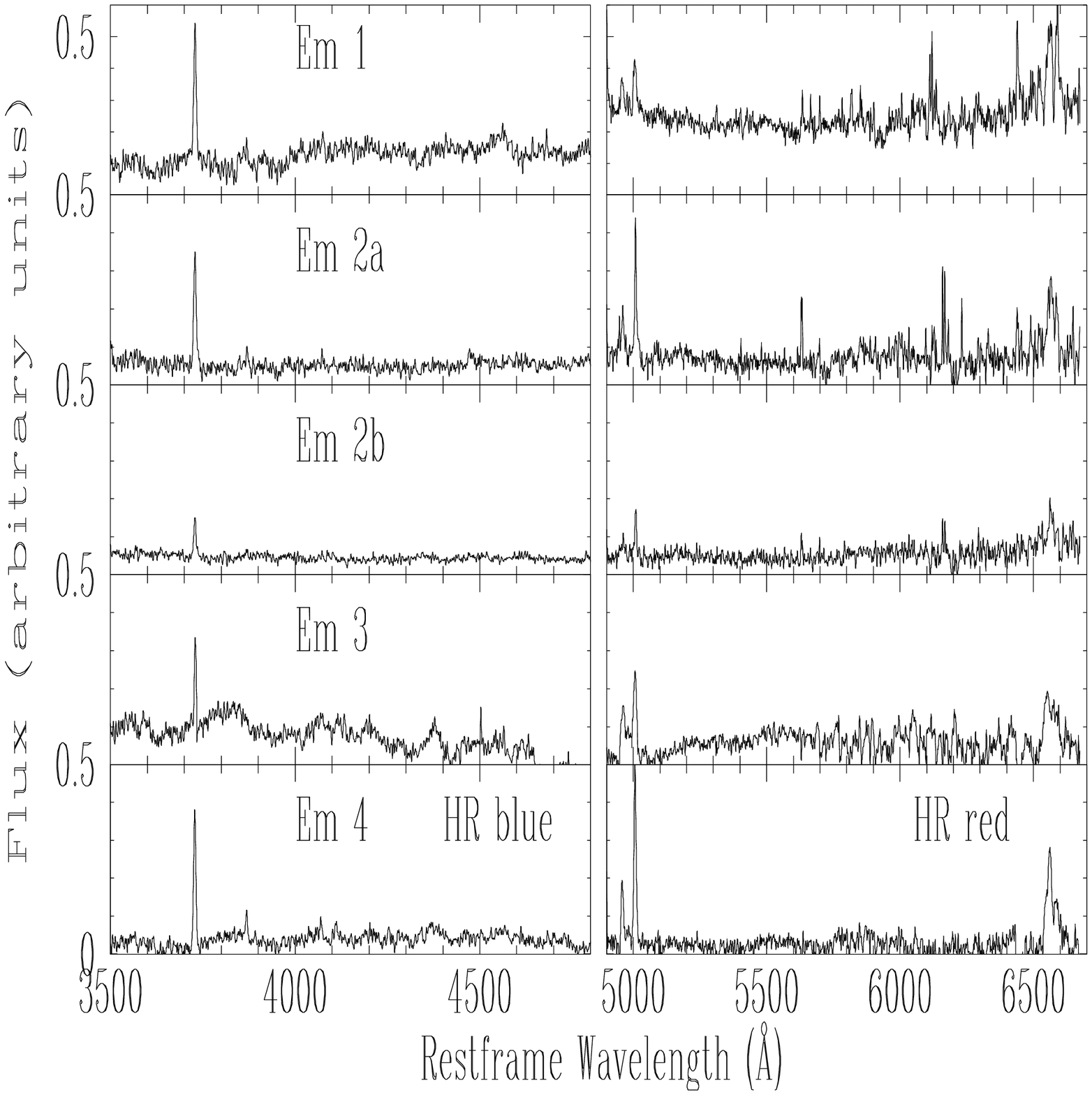}
\caption{Spectra corresponding to the regions labelled Em1 to Em4 in Fig.~\ref{zones},
all dominated by ionized gas.}
\label{Em}
\end{figure*}

Finally, some of the slits  used for these observations  were placed on galaxies
close to  the line of  sight to  \obj.   One of  them is  galaxy G1 of
Fig.~\ref{preim},   for    which     we  obtain     the   spectrum shown in
Fig.~\ref{neigh}.  The redshift of this   late-type galaxy is close to that  of
the quasar, $z=0.2866$. The projected linear distance  between G1 and the quasar
is 44 kpc. No other related galaxy was found in these and  the previous FORS1 spectra (\cite{letawe07}).


\section{Integral field (3D) spectroscopy}

\subsection{Subtraction of the point sources}
\label{subps}
Integral field spectroscopy  (IFS) allows investigation of the  immediate
surroundings of the QSO, provided its own spectrum can  be subtracted from the
data.  Because of the  lack of isolated stars   in the VIMOS field  of
view, and because of the elongation of the PSF, deconvolution of the data,
as done with slit spectra, is not possible. Instead, we use
another approach,  taking advantage of the 3D  information  we have at
our disposal.

First, we select  a reference spectrum at  a central position for each
point  source,   i.e.,    the  QSO   and  the    foreground   star  of
Fig.~\ref{imgHST}.    The  two spectra are  used    as templates to be
removed from the other spaxels, after adequate weighting.  As there is
a clear  variation  of the  grism+detector response  function  accross
the array  of  fibers,  the weighting   is allowed  to  vary linearly with
wavelength, for each spaxel.

The relative weighting  of the QSO and  of  the stellar templates  are
considered as satisfactory  when there  is no trace  left, in  a given
spaxel, of the  Balmer broad emission lines  from  the QSO and of  the
(broad) absorption  lines  for  the  star.   Visual inspection    of the
residuals is carried out  for every  spaxel contaminated by a point source, allowing us to make sure that the result of the subtraction does not show
signs of over- or undersubtraction, at  any wavelength. This procedure
leaves us  with a data  cube containing only  the companion galaxy and
any other extended objects such as regions of ionized gas. 

The method described  above relies on  the important assumption  that
the QSO template  actually contains  only light  from the  QSO, which
might not be the case. Our template probably  contains a small but non-zero
contribution  of (narrow)  gas  emission  lines not belonging  to  the QSO
itself.   We may  therefore  systematically oversubtract  the emission
lines in each spaxel, in particular for the  ``blob'', which lies angularly close to
the   QSO.   Moreover,   the subtraction
significantly increases the  noise   in the  regions where the   QSO
or  the star dominate the flux, i.e. close to the centers of their PSFs.

The adopted procedure for point source subtraction being based on visual inspection of the removal of spectral point source characteristics, the error bars are difficult to quantify precisely. However, as we aim to subtract the broad components with no condition on narrow ones, the presence of narrow emission lines in the subtracted spectra is reliable, and might just be slightly underestimated by the possible subtraction of part of the blob spectrum along with the quasar template. The continuum slope might also be affected by a wrong weighting of the template spectra, so we did not use these slopes to determine any properties of the emitting regions.

\subsection{Extracted spectra}
\label{exspec}

The   result   of  the    QSO+star  subtraction    is   displayed   in
Fig.~\ref{zones}.  In this figure, we also divide the field of view into
different regions where we integrate  the signal.  Three main areas are 
defined and then subdivided into smaller areas: the region of the companion
galaxy, the more central region of the ``blob'', and several regions further away from the QSO,
 which also turn out to contain ionized gas.

\subsubsection{The companion galaxy}

The companion galaxy can be subdivided into three areas which we
name {\it Gal1} to {\it Gal3} (Fig.~\ref{zones}). The corresponding 1D spectra
are displayed in Fig.~\ref{gal}.  While the spectrum of region 
 {\it Gal1} is  dominated by  young
stars  (important  continuum, weak  [OIII]  lines,
broad Hydrogen absorptions),  region  {\it Gal2} is a mix of star and gas,
and region  {\it Gal3} is almost  exclusively made of  ionized gas.  As  the
smearing of light is of consequence in the VIMOS data, we deliberately integrated the signal over large regions, which have about the length-scale 
of the seeing. The matching between the regions and the HST/ACS broad-band 
image is shown in Fig.~\ref{zones}.   Clearly, regions of  highly ionized gas 
extend  beyond the broad-band image of the galaxy: there is basically no light in the HST/ACS image of the {\it Gal3} region, as well as no continuum in its spectrum.

\subsubsection{The blob}

The region of the blob is probably the most difficult to interpret on the VIMOS
data, as it is located very close to the reference fiber taken to model our QSO 
template spectrum. Although the subtraction process may bias the spectrum
of the blob (essentially by removing part of it together with the QSO spectrum), the present results support the M05 findings that the 
blob is made only of gas and shows no trace of a stellar continuum.
The 1D spectrum of this region is shown in Fig.~\ref{blob}.

Note that even if our subtraction procedure dimmed the continuum 
of the blob, it has also dimmed the emission lines. The fact that the latter are still very bright after subtraction of the QSO spectrum shows that we  have not completely removed the signal.

\subsubsection{Emission line clouds}

The four regions labelled {\it Em1} to {\it Em4} in Fig.~\ref{zones} 
are dominated by emission lines. Their integrated 1D spectra are shown in Fig.~\ref{Em}. 
All emission lines are real, as no appropriate weighting  of the subtracted  nuclear
spectra can  make them disappear, even when trying to oversubtract the templates. 
At the depth of the observations, only {\it  Em1}  contains both gas and
stars, as can be seen from the non-negligible 4000\AA\ break. This region might
be an extension of the companion galaxy. The three  other
regions are compatible with gas emission  only. The weird continuum of
{\it Em 3} is caused by its proximity to  the center of the foreground
star, these  spaxels  being more  affected  by the PSF variation  with
position  and wavelength. Nothing can be  seen at the position of {\it
Em2}, {\it Em3}  and {\it Em4} in  the HST/ACS image (Fig.~\ref{zones},
left). {\it Em3} probably   corresponds  to the faint    gaseous
emission R2 in the VLT MXU spectra of Fig.\ref{deco3}.

\subsubsection{A 2D diagnostic diagram}

\begin{figure*}
\centering
\includegraphics[width=15.cm]{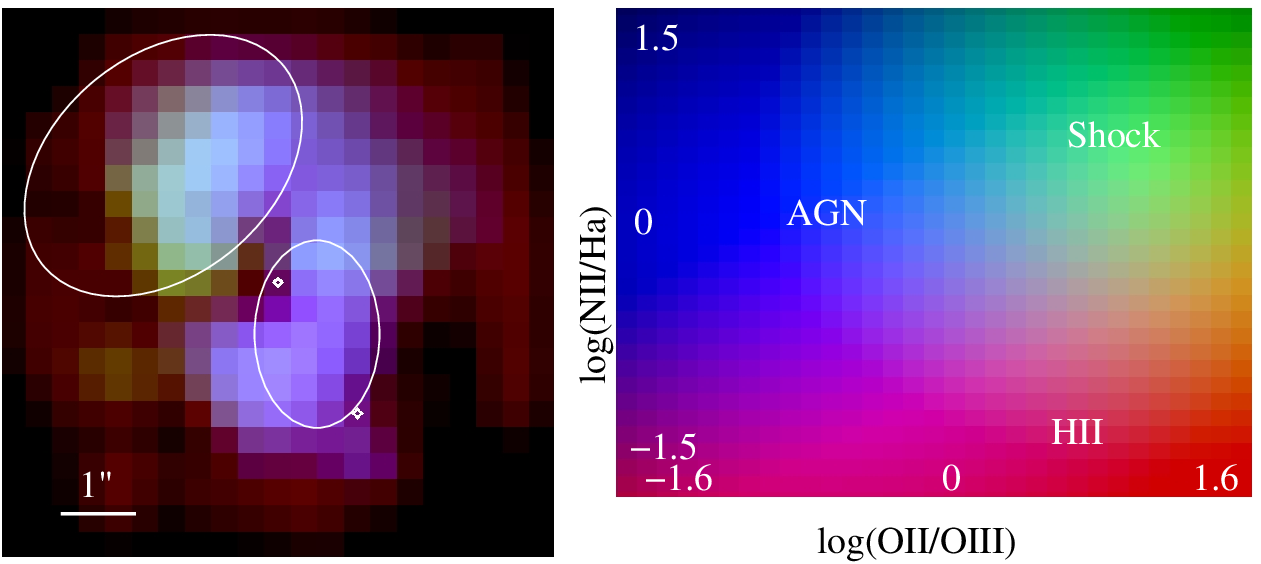}
\caption{Color-coded diagnostic diagram constructed from the VIMOS 
emission line images after subtraction of the QSO and of the nearby star, 
as in the right panel of Fig.~\ref{zones}. The positions of the QSO and of the 
nearby star are indicated by white diamonds, while the companion galaxy 
and the ``blob'' are indicated by white ellipses. 
The color code in the image is defined in the right panel.  The field of view 
is  $7\times 7\arcsec$, i.e., a zoom  on the central part of the VIMOS field.  
The image has been slightly smoothed, and weighted at each position by the corresponding emission intensity.}
\label{dd}
\end{figure*}

Diagnostic diagrams such as the ones introduced by \cite{veilleux} and
\cite{baldwin} allow identification of the nature of the source of ionization 
in a gas cloud. These diagrams involve emission line ratios sensitive to
the hardness of the ionizing flux. Our VIMOS data allow us to map some of these
ratios accross the field of \obj.


We  show in Fig.~\ref{dd} a color-coded 2D diagnostic diagram 
for the regions in the vicinity of \obj.   The 3-color scale in the right panel gives the ionization source as a function of color.
The   three main sources of ionization are labeled in the figure as ``AGN''  for ionization by  strong power-law source (here the QSO), ``HII'' when the ionization is due to stars,  and ``shock'' for shock-induced ionization. 

In order to be accurate, the line ratio must be corrected for absorption by dust.
The lack  of an H$\beta$ line in these spectra makes it  hard to carry out this 
correction and hampers the use of the line ratio proposed by \cite{veilleux}. However,  our diagnostic   diagram  is constructed using the [NII]/H$\alpha$ and[OII]/[OIII]  line ratios, as presented in \cite{baldwin}.  With these ratios, any reddening leaves the  [NII]/H$\alpha$  unchanged, as the two lines are almost at the same wavelength, but decreases the [OII]/[OIII] ratio:  [OII] being at shorter wavelength than [OIII], it is much more affected by dust absorption.
. 

The vast majority   of spaxels are found   in the AGN-ionized  region (blue
dominant). They  concentrate in the blob, and in the regions {\it Gal2/3} and
{\it Em3/4}. The Southern  part of the system ({\it Gal1/Em1})
is ionized by shocks and stars. Following Balmer decrements evaluated on FORS1 slit spectra (M05), the companion galaxy is the only place where dust reddening is non negligible. However, a correction for this extinction, increasing  [OII]/[OIII], would not affect the dominance of the shock-induced ionization. 

 As the ionization source of the blob is clearly the quasar light, the possible oversubtraction of a part of the blob (narrow component) along with the quasar template spectrum (as explained in Sect.~\ref{subps}) possibly lowers the hard ionization across the entire system. The dominance of AGN-caused ionization measured in the whole system may thus be even stronger than suggested by Fig.~\ref{dd}.

The only regions where the  ionization by star  dominates, are the region {\it
Em2b}  and the outer edge   of region {\it Gal1}, where the signal is very 
weak.  Even if we find almost no  continuum in the weak  {\it Em2} region, 
this might indicate the presence of young stars even outside the companion galaxy.
The distribution of the green shock-induced ionization regions in Fig.~\ref{dd}, on each side of the QSO, suggests that the stars were formed during a violent event, probably triggered by the QSO (or together with the QSO) and not in a global, continuous star formation in the companion galaxy. 

Finally, note that the spectra of the different  {\it Em} zones are
very faint, and that  H$\alpha$ and [NII]  are not well separated in
those regions (see Fig.~\ref{Em}), resulting in large uncertainties on
the measured ratios. Still, the measured [NII]/H$\alpha$ would have to be
wrong by an order of magnitude to change our main conclusion that the 
gas is ionized by a hard radiation field, probably that of the QSO.


\subsection{Emission line images}

\begin{figure}
\centering
\includegraphics[width=8.5cm]{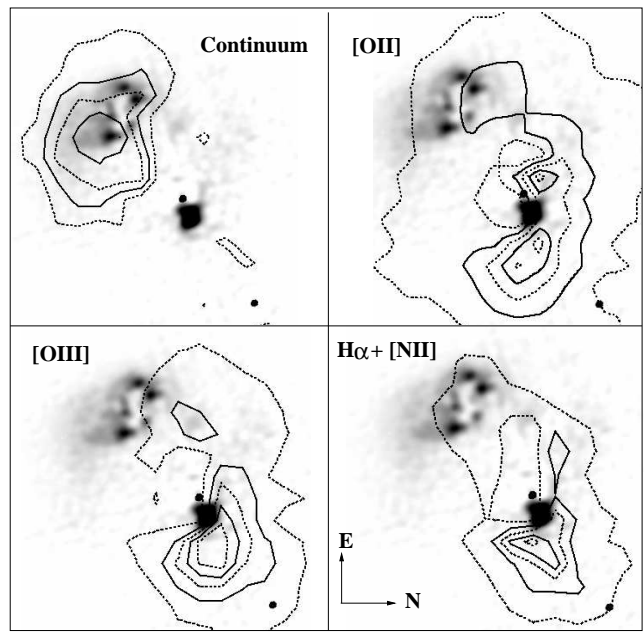}
\caption{
Contours plot for the continuum VIMOS image (see text) and for 
different emission lines. The contours are shown in overlay of the HST/ACS
image, after subtraction of the QSO and of the nearby star. We alternate solid and dotted contours for clarity of display. The steps between contours are every 20\% of the maximum intensity of each image.}
\label{raiesem}
\end{figure}

We use our VIMOS data to produce narrow-band images
of \obj, centered on several emission lines:   [OII]3727\AA,  [OIII]4959+5007\AA,   and
H$\alpha$6563\AA\, + [NII]6583\AA. For  each line, we subtract an image
in the continuum   around     the emission line,  after    appropriate
normalization.  We also show an  image  of the continuum between [OII]
and  H$\beta$,  a wavelength range  devoid   of any emission lines. 
We overlay on Fig.~\ref{raiesem} the contours of emission lines images
on the deconvolved HST/ACS data, where point sources are indicated by
dots. We set  to zero the  noisy pattern near the position of the
QSO, because of  the   large uncertainty due to the removal of its spectrum at this location.

We observe a slight shift between the position of the  blob
in the HST image and in  the VIMOS contour plots. The reason for this   
is twofold: first, as we take a reference spectrum at the central  position for PSF subtraction,
we remove part of  the underlying extended component (i.e.  the blob) together with  the
quasar light at  those positions. Second,  the spatial resolution of
VIMOS is much lower than that of ACS, so small emission regions in the HST
data are spread on  much larger areas in the VIMOS data. One spaxel in
the VIMOS data is about the size of the blob. 

The most striking differences between the spatial distribution in the 
various spectral bands are given below:

\begin{itemize}

\item The continuum light nicely follows the ACS image of the companion 
galaxy.  No continuum  is found at the  position  of the blob. Only small and faint
structures are seen in the vicinity of the QSO and are all compatible with 
the noise. 

\item The [OII] emission is present in the companion  galaxy, the  blob, 
and extends all around \obj. This
line arises both in star forming  regions and in  gas clouds ionized by
AGN light. Its extended distribution indicates that the system  seems globally embedded in
gas. Note that the effect is not due to an arbitrary choice of the intensity scales:
although the [OII] envelope is fainter than the [OIII] emission (see, e.g.,  
Fig.~\ref{Em}), it extends much further away from the QSO. 

\item The [OIII] doublet needs harder ionizing photons than [OII], such as
ionization by an AGN. In Fig.\ref{raiesem},  [OIII] is essentially present 
in the blob, in the  Northern part of the companion galaxy, as well as close to the  
position of the foreground  star.

\item The H$\alpha$ distribution is more extended than [OIII] in the companion galaxy, 
probably because of ongoing star formation. The blob is also visible in this line.

\end{itemize}


\subsection{Velocity maps}

\begin{figure}
\centering
\includegraphics[width=8.cm]{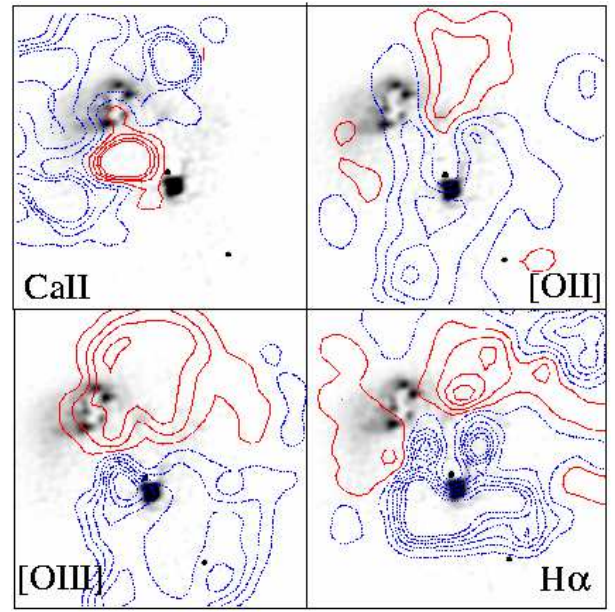}
\caption{Velocity maps for the stellar and gas components, overlaid on the HST/ACS deconvolved image.  The velocity contours are relative to the companion galaxy systemic velocity. The blue dashed contours   are negative, and the red solid contours 
are   positive. Top left: CaII absorption line velocity contours, from --200  to +200$\kms$ with steps  of 55$\kms$. Top right: [OII]  emission line velocities, where the contours are drawn from --190 to +160$\kms$ by  steps of 50$\kms$. Bottom left: the [OIII] emission line, with contours  from --290 to +160$\kms$ with    steps of 50$\kms$. Bottom right: H$\alpha$ emission,  from --290  to +160$\kms$ with steps  of 60$\kms$. See text for interpretations.} 
\label{velo} 
\end{figure}

The velocity maps  of Fig.~\ref{velo} are constructed 
by fitting a Gaussian profile to the emission lines in  each spaxel. The velocities
are then given relative to the companion galaxy average velocity, as estimated from the {\it Gal1} emission lines. The velocity contours obtained in this way are then slightly
smoothed.  With the setting of the VIMOS observations, we can map
both the gas and stellar velocity fields.

The best lines to trace the gas in our observations are [OII] and
[OIII]. The H$\alpha$ velocity map is noisier because of the high nucleus-to-host ratio in this line. Spaxels in the immediate surrounding of the quasar position are masked. The  velocities derived from these lines range from --400 to +150$\kms$, with respect to the companion galaxy (or from about --300 to +300$\kms$, with respect to the QSO). The gas can be traced very far away from the
QSO, even where no light is seen in the HST image, as already seen in Fig.~\ref{raiesem}. Even with discrepancies between [OII] and H$\alpha$ velocity fields, the global trend is to show positive velocities  on both sides of the galaxy, and negative elsewhere. The [OIII] velocity field is slightly different, globally positive on  the whole upper part of the field and negative West of the quasar. This reinforces the hypothesis of a highly ionized gas cloud traced by [OIII], detached from the galaxy to which [OII] and H$\alpha$ are more related.

The extraction of the stellar velocity field, from the CaII absorption lines,
is  restricted to the  companion  galaxy. Our measurements are presented 
in Fig.~\ref{velo} (top left), ranging from --150 to 300$\kms$. The difference with the gas velocity field 
is striking, with  almost reverse velocities, positive between the position of the galaxy and the quasar, and negative elsewhere. Taking into account the elongation of the VIMOS PSF, the rotation axis probably matches the major axis of the galaxy.

Clearly, gas and stellar dynamics  in the companion 
galaxy are not   related.  Either the gas and the stars from which we measure the
dynamics are located in physically different regions seen at the same position
on the plane of the sky due to projection effects, or the  gas and the stellar 
spectral features are formed in the same location. The first hypothesis 
would fit with the idea that \obj\ is embedded in a large cloud of gas ionized by
the QSO radiation field at very large distances. The second hypothesis would 
rather indicate a violent shock between galaxies with the possible disruption of the 
host galaxy. 

\section{Ionized gas and radio jets}

Ionized gas is present in a large volume around \obj, as shown both by the slit and IFU
observations.  Two questions thus arise: (1) where
does the gas  come  from ?  (2) how  can it be so highly ionized 
as far away as 30 kpc from the QSO  ?


Both questions may find an  answer in the  context of a collision
between the companion galaxy and  the putative quasar host, dispersing
the gas and allowing the hard radiation field to reach large distances from
the  nucleus because of the very irregular  gas (and possibly dust) 
distribution around the QSO. 

The immediate surroundings of the QSO obviously  contain some gas (such as the blob), but are also sufficiently transparent to let the QSO radiation field escape to large distances.

The detection of an extended gas distribution in the [OII] 3727\AA\, line tends to weaken the hypothesis that the QSO host galaxy might remain undetected because of strong absorption by dust. Indeed, if the whole host was embedded in (or hidden behind) a huge dust cloud, it is difficult to understand how radiation at such short wavelengths might be detected throughout the system.

 Deep infrared observations, 
aimed at testing this hypothesis are currently under analysis. 
However the presence of dust cannot be ruled out for the moment, and in this case the radiation of the quasar might evaporate the dust in some loci, producing patchy and highly ionized gas clouds.

\begin{figure}
\centering
\includegraphics[width=9.cm]{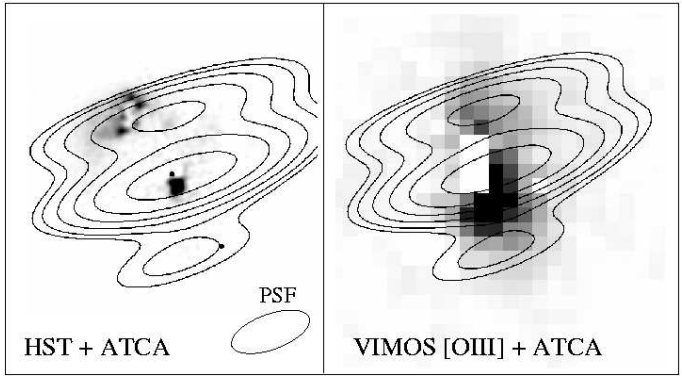}
\caption{Radio map  obtained from the ATCA observations of \cite{klam07},  
overlaid on the deconvolved HST image (left) and on the VIMOS [OIII] image (right). 
Note the strong correlation between the radio structures and the [OIII] image and
the misalignment with respect to the companion galaxy.}
 \label{radio}
 \end{figure}

Whatever the origin  of the gas patches, a  possible source of
ionization in  addition  to the quasar  radiation  itself  is a  radio
jet/cloud  interaction,  as already proposed by \cite{klam07}. In Fig.~\ref{radio}   we reproduce their  6208 MHz ATCA radio data,  overlaid on the HST  image
and on the VIMOS [OIII]  image. If dust was present on the path of the radio jets and lobes, appropriate correction for reddening might shift the ionization source  for these regions from AGN to shock-induced ionization, these shocks then being caused by the jet.   As seen from Fig.~\ref{radio}, there is a clear alignment
between  the radio lobes and  the [OIII] emission.  The 
alignment  is  found only between the radio data and the [OIII]  data, 
rather  than with  [OII] and H$\alpha$, suggesting 
that   the  radio jets are spatially   related  to high  level
ionization clouds,  and not to  star forming regions. However, as the flux used in the  radio-to-infrared flux ratio estimated  by \cite{klam07} concerns only the central peak of radio emission, this leaves their conclusions basically unchanged.

In addition, the strong spatial correlation between the  radio jet and the [OIII] emission 
suggests that the jet is either the source of ionization or an efficient cleaner  of  the   surroundings of the QSO, allowing its 
radiation  to  escape from  the central  region and to ionize much  more distant gas
clouds.

\section{Discussion}

Our new VLT 2D and 3D spectroscopic observations provide further evidence that the enigmatic QSO HE0450-2958 lies in a strongly perturbed environment.  
\begin{enumerate}
\item The putative host remains undetected, the companion galaxy is strongly distorted, with a gas velocity field apparently unrelated to the stellar one.  
\item  The whole system seems embedded in ionized gas, as best shown by the extended distribution of the [O II] emission.  
\item  Radio emission shows a peak at the position of the QSO as well as two roughly symmetrical lobes, one of them being located close to the companion galaxy.  
\item  Several clouds of highly ionized gas (those with the highest [O III]/[O II] ratio) are found along the radio axis, extending from the immediate vicinity of the QSO (i.e. the ``blob") up to more than 30 kpc. 
\item  These gas clouds appear to be ionized either by the QSO radiation itself or by shocks associated with jets.
\item  This long range ionization suggests that the surroundings of the QSO are basically transparent to UV radiation, at least in two directions (along the radio axis and towards us - we recall that the blob spectrum shows no reddening - M05).
\item  The facts that the radio lobe and long range ionization are stronger in the S-W direction (i.e. towards the companion galaxy) than in the opposite direction might be due to the fact that the gas cloud immediately N-E of the QSO (the blob) absorbs most of the energy emitted in that direction. It may also be explained by a larger amount of gas in the direction of the companion galaxy than on the other side of the QSO/blob.
\item Several of these observations may find an explanation in the context of a collision which would have dispersed matter in a large volume around the QSO and companion galaxy.
\end{enumerate}

Since HE0450-2958 is a strong infrared emitter ( IRAS point source catalog, \cite{degrijp}; \cite{low}), dust should be present in its surroundings,   but the spatial resolution of the IRAS observations used for this identification does not allow us to locate precisely its origin.  Kim et al. (2007) explained why this strong infrared emission cannot be caused by star formation in the companion galaxy: If the IRAS measured infrared fluxes  were to be associated with star formation, they would imply a star formation rate of almost a thousand M$_\odot$/yr, while the strength of the [OII]3727$\lambda$ line in the optical FORS spectra of the companion galaxy only gives a maximum of $\sim$ 10 M$_\odot$/yr. Accordingly, the strong infrared emission is thus most probably not associated with the companion galaxy, but rather with the quasar itself. A simple explanation would be a dust torus around the accretion disk.  The fact that the blob is unreddened suggests that either this torus is quite compact (at most a few hundred parsecs in order not to cover the blob) or the blob lies in front of it.  The dust torus should also be inclined at an intermediate angle with respect to the line-of-sight.  Indeed, an edge-on torus can be ruled out as it would mask the QSO.  Similarly, a nearly face-on torus would hardly be compatible with the fact that the QSO UV radiation reaches projected distances as large as 30 kpc.

The fact that the UV radiation field of the QSO, after having left its immediate surroundings, can escape freely to large distances suggests that, at the kpc scale, the QSO neighbourhood is basically transparent in several directions. Even if we have no evidence for the complete absence of a host galaxy, this transparency is much easier to explain if the QSO does not lie in a massive host.  Indeed, if such a galaxy was present in an environment so strongly perturbed by gravitational interactions, one would expect starburst regions with large amounts of gas and dust, as can be seen in the companion galaxy.  Such a medium would not likely be transparent enough to explain the long range ionization.  The present observations might thus reinforce the hypothesis of an undermassive host, if any at all. 

 On the other hand, the analysis of radio by \cite{klam07} seems to point towards some star formation around the quasar, leaving the possibility of a dust enshrouded host galaxy, explaining its non-detection in the optical, a dust medium which could have been swept in several directions by the radio jets, allowing the observed long range ionization by the powerful quasar.

\begin{acknowledgements}
G.L. is a teaching assistant supported by the  University of Liege. GL
and PM acknowledge  support from Prodex  90195 (ESA/PPS Science Policy, Belgium). FC is supported by the Swiss National Science Foundation (SNSF).
\end{acknowledgements}

\end{document}